\documentstyle[twocolumn,prl,epsfig,aps]{revtex}
\begin{document}
\preprint{LPTENS-97/18}

\noindent
{\bf Cugliandolo and Iguain reply}:
In~\cite{LJL} we stated precisely 
that the aim of our study was to rule out
non-resonant hole-burning (NSHB) experiments as proofs for the existence of 
{\it spatially} heterogeneous regions in glasses.
The results of NSHB experiments have been often cited as
evidence for the existence of 
{\it spatially distinguishable, dynamical, regions} in the samples. 
The list of references is long, see {\it e.g.} \cite{review}.
Agreement about the limitation of NSHB to determine the existence of 
{\it spatial} heterogeneities is now reached~\cite{Dieze,Cham}.

\begin{figure}
\centerline{\hbox{
   \epsfig{figure=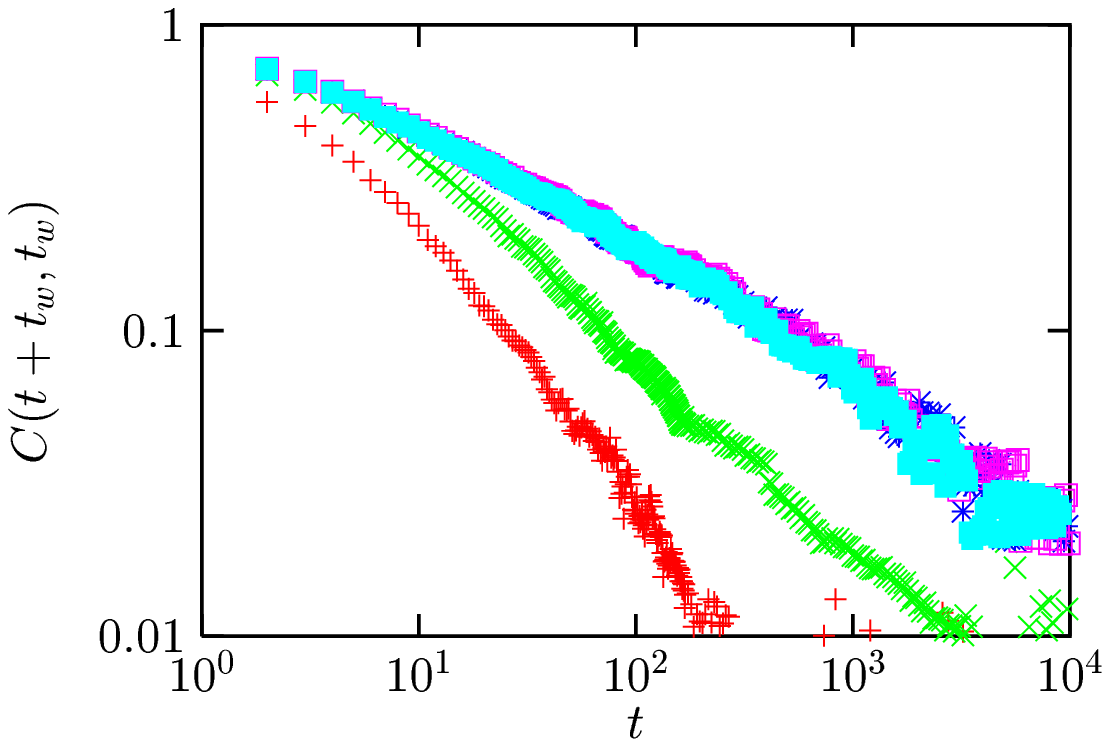,width=7cm}}
 }
\caption{Auto - correlation of the SK mean-field spin- glass model 
in the absence of an external field
for $t_w=1, 100,1000, 25000, 50000$ MC steps from bottom to top,
the last three curves collapse.
}
\label{fig1}
\centerline{\hbox{
   \epsfig{figure=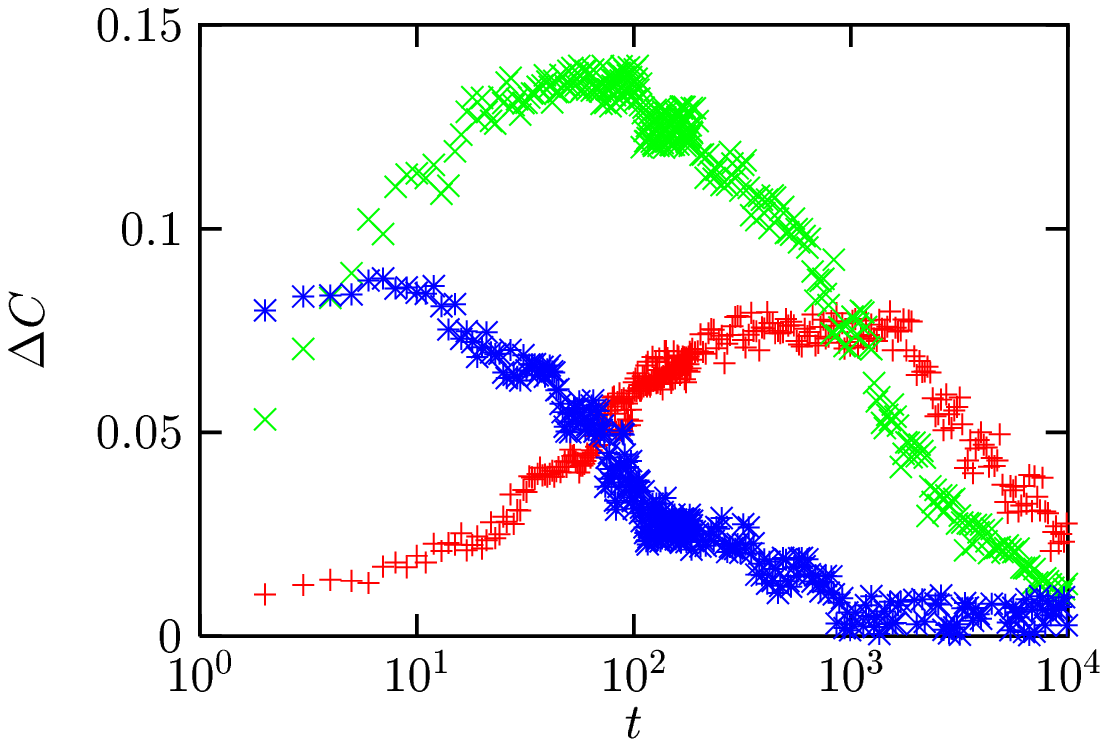,width=7cm}}
 }
\caption{The distortion $\Delta C$ after a single pump oscillation.
$h_{\sc f}=1, t_w=25000$ MC steps. The symbols $\star$ (blue), $\times $ 
(green) and $+$ (red) correspond to 
$\omega=1.5,0.1,0.005$, respectively.
}
\label{fig2}
\end{figure}   
Diezemann and B{\"o}hmer point out, as we have also done in our article, 
that our measurements were done out of equilibrium. At least two of
the NSHB experiments have been done below $T_g$, where equilibration 
is impossible~\cite{belowTg,chamberlin} (see \cite{Nordblad}
where it is proven that even a field-cooled spin-glass may show
nonequilibrium relaxation). Aging effects and the 
relaxation due to an applied ac pumping field
have not been explicitly disentangled in these references. 

The question as to whether the NSHB protocol applied to an equilibrated  
fully connected model (without notion of
space), can reproduce the results of these experiments
is very interesting indeed and deserves careful inspection.
We take this opportunity to present results from a Montecarlo simulation 
of the Sherrington-Kirkaptrick (SK) mean-field spin-glass model 
with $N=3000$ spins 
at a temperature  $T=T_g-\epsilon=0.99$. 
Figure~\ref{fig1} shows one of the standard checks of equilibration 
that consists in monitoring the autocorrelation 
function  $C(t+t_w,t_w)$  for different waiting-times
$t_w$ and observing when it starts being stationary.
This happens for $t_{eq} \approx 25000$ MC steps.
Other, more subtle methods can also
be used~\cite{peter} and confirm  equilibration
at this waiting time. In Fig.~\ref{fig2} we apply the NSHB protocol
after $t_w=25000$ MC steps with one oscillation of an 
ac-field with  $h_{\sc f}=1$ and $\omega=1.5,0.1,0.005$; $t_r=0$.
This figure can be compared, for instance, to Fig.~2 in
Ref.~\cite{chamberlin}
and, as far as we can judge, all qualitative features, including 
the shift in the base of the spectral hole mentioned
in \cite{Cham}, are reproduced.

The effect of several cycles has already been discussed in~\cite{LJL}.
A thorough discussion of where in parameter space 
aging effects disappear under a permanent ac-field 
in this model appeared in~\cite{LJLL}. Once the stationary regime (in 
stroboscopic time) is reached, the spectral modification will obviously 
saturate. 

 In the $p$-spin and SK models 
it is not possible to identify localized domains and, however, the response 
to NSHB looks pretty much the same as in real systems.  
\vspace{.7cm}               

\noindent
\small{
Leticia F. Cugliandolo$^{1,2}$ and 
Jos{\'e} Luis Iguain$^{2}$ \\
$^1$Laboratoire de Physique Th{\'e}orique,
{\'E}cole Normale Sup{\'e}rieure,
24 rue Lhomond, 75231 Paris Cedex 05, France \\
$^2$Laboratoire de Physique Th{\'e}orique  et Hautes Energies, Jussieu, 
5{\`e}me {\'e}tage,  Tour 24, 4 Place Jussieu, 75005 Paris France
}

\vspace{.5cm}   
\noindent
Received 2001


\end{document}